# Designing a Multi-petabyte Database for LSST

Jacek Becla[1a], Andrew Hanushevsky[a], Sergei Nikolaev[b], Ghaleb Abdulla[b],
Alex Szalay[c], Maria Nieto-Santisteban[c], Ani Thakar[c], Jim Gray[d]

[a] Stanford Linear Accelerator Center, 2575 Sand Hill Road, Menlo Park, CA, 94085, USA
[b] Lawrence Livermore National Laboratory, 7000 East Avenue, Livermore, CA, 94550, USA
[c] Johns Hopkins University, 3400 N. Charles Street, Baltimore, MD, 21218, USA
[d] Microsoft Research, 455 Market Street, 16th floor, San Francisco, CA, 94105, USA

## ABSTRACT

The 3.2 giga-pixel LSST camera will produce approximately half a petabyte of archive images every month. These data need to be reduced in under a minute to produce real-time transient alerts, and then added to the cumulative catalog for further analysis. The catalog is expected to grow about three hundred terabytes per year.

The data volume, the real-time transient alerting requirements of the LSST, and its spatio-temporal aspects require innovative techniques to build an efficient data access system at reasonable cost. As currently envisioned, the system will rely on a database for catalogs and metadata. Several database systems are being evaluated to understand how they perform at these data rates, data volumes, and access patterns.

This paper describes the LSST requirements, the challenges they impose, the data access philosophy, results to date from evaluating available database technologies against LSST requirements, and the proposed database architecture to meet the data challenges.

## 1. INTRODUCTION

The LSST will start producing data in 2012. It is expected to produce over seven petabytes of image and catalog data per year. The image data is reduced by processing pipelines that recognize and characterize objects. The catalogs for these objects and their metadata will grow at an estimated 300 terabytes per year. They will reach multi-petabytes during the lifetime of the survey. Today, most "very large" databases are still measured in tens of terabytes, with only a few reaching hundreds of terabytes[2]. While it is likely many more will reach peta-scale by 2012, designing a database system for managing multi-petabytes inevitably poses significant challenges.

Existing methods are not applicable in the peta-scale. A single sequential scan through a megabyte takes less than a second, but at 10 MB/s it would take over a year to complete such a scan through a petabyte. Even well-known, efficient search approaches, such as indexes do not easily scale to the LSST levels. Using traditional indexes to make compact extracts of the data will be problematic; the size of an index for a petabyte of data is likely to reach many terabytes, which means a single sequential scan at 10MB/sec would take more than 2 days. With this in mind, and with the expertise in building a production ODBMS[3]-based petabyte database for BaBar - a High Energy Experiment [1], and multi-terabyte RDBMS[4]-based database for Sloan Digital Sky Survey (SDSS) [2,3] we recently embarked on designing the LSST database. This paper discusses the current LSST Database design and rationale. Chapter 2 describes LSST-specific requirements and challenges, Chapter 3 outlines the planned architecture of the LSST database, Chapter 4 covers data ingest issues, Chapter 5 covers our approach to designing and building the system, and Chapter 6 summarizes the discussion.

---

[1] becla@slac.stanford.edu
[2] As of this writing, we are aware of just two databases that reached a petabyte. One is based on object oriented database, and was designed and managed by some of the authors of this paper [1], the other, a Sprint/HP SQL system, is a 1.1 PB system reported in page 9 http://www.wintercorp.com/VLDB/2005_TopTen_Survey/2005TopTenWinners.pdf.
[3] ODBMS – Object Oriented Data Base Management System
[4] RDBMS – Relational Data Base Management System

## 2. REQUIREMENTS AND CHALLENGES

The main LSST Database challenges include large data volumes, high availability, real-time transient alerts, spatio-temporal indexing and search, public access, data reproducibility, and data provenance. Details for each listed challenge with related requirements are given below.

The database system has to efficiently manage <u>large volumes</u> of data: approximately three terabytes of new image data will be added to the database every 24 hours. As the data volume grows, it will be harder and harder to re-index petabytes of data, or track changes to billions of objects. Due to high cost of storage, the bulky data objects that can easily be rebuilt will not be persisted[5], instead detailed <u>provenance</u> will be recorded in database so that these derived data products can be recomputed as needed..

With large volumes come <u>high availability</u> challenges: archiving many petabytes of data will require tens of thousands of disks, even if disks are terabyte scale. While the mean time to failure (MTTF) of one disk is several decades, the MTTF of tens of thousands of disks is measured in hours. Daily disk failures must be planned for, and recovery must not affect the availability of the database. Commonly known techniques such as data replication have to be carefully applied, as they tend to be very expensive on this scale.

LSST will generate <u>real-time transient alerts</u>; the planned latency should not exceed 60 sec total, including database operations. As a result, data processing and publication must be an almost real-time system.

LSST is going to explore both <u>spatial and temporal</u> aspects of the data. This will require sophisticated techniques to efficiently index data for spatial and temporal queries across multiple observations. Past surveys mostly focused either on spatial or temporal aspects, but rarely both.

All LSST data products will be made <u>public</u> with minimal delay, and should be easily accessible. The system must provide easy and quick query access to a wide range of users: professional and amateur astronomers, students and the general public. Also, it should be possible to cross-correlate LSST data with data from other astronomical surveys.

Results published based on released data must be <u>reproducible</u>. That means that released data may never change and the system has to track software configurations and data versions.

## 3. PLANNED ARCHITECTURE

Systems for managing very large data sets are complicated by nature. The architecture of the LSST Database is no exception, and while we are striving to keep it as simple as possible and not over-design it, LSST is poised to use advanced database features like partitioning, distribution and replication as discussed here. Further details regarding the techniques and features expected to be used in LSST are discussed below.

### 3.1 Hybrid approach

One of the key LSST Data Management design concepts is the "hybrid" approach: LSST will keep all images in flat files, and "everything else" will be managed by a database. The database will be used to:

a) store and manage all data derived from the images, called *Catalog Data*, and

b) track the projects' metadata, including data versions, provenance and hardware configurations.

The main reason why we decided against keeping images inside the database is to maintain portability and database-independence as well as scalability. Furthermore, we saw no apparent benefits of having the images inside a database, and given the volume of image data (several times the catalog and metadata), the potential to aggravate scalability issues. Portability is particularly important for the raw images. Since they will almost certainly be used long after 2022 -- the LSST's planned shutdown, it is essential to store them using as portable technology as possible. Each image file will carry its own metadata as part of its FITS header [9]. Image files will be immutable, that is once written they will never change. Raw images will never be deleted. Any updates to the image metadata will be tracked by the database and re-generated on-the-fly as needed.

Image files at the file system level will be managed using either a parallel file system like Lustre [7] or Ibrix [8], or a file management system like xrootd [4] or SRB [6].

---

[5] If we retained all level 2 data products, the database would grow about twice as fast as current estimates.

The hybrid approach and the co-existence of two storage technologies (files and databases) will require glue between them. By storing "reference" to image files, the database will act as that glue. The proposed technique involves keeping logical paths to files in the database, and mapping these paths to the corresponding physical paths on the fly. These logical paths can be reconstructed from the files e.g. by scanning their headers. This approach decouples file management from the database, and allows the file system to freely relocate files without any changes to the database. This approach is similar to that used by BaBar and SDSS.

### 3.2 Distributed, tiered architecture

Another equally important design concept is a highly-distributed, tiered architecture. Acquired data will be transported from the mountain summit (i.e., the location of the telescope) to the Base Facility (i.e., a local data processing center nearby the LSST telescope,) to the Archive Center in the United States.. Once processed, the catalog and image data will be replicated at multiple sites (tiers) [12]. The Base Facility processing will generate real-time transient alerts. The Archive Center will play multiple roles including bulk processing, long term archive, replica management, and providing efficient access to data for other tiers and for scientists to do analysis. From the Archive Center, the data are distributed/replicated to multiple Data Access Centers and Tier-n End User Sites for community access. We expect that the Archive Center and other tier 1 centers will be active participants in the Virtual Observatory linking all the world's science archives [10]. The database will play a vital role at all these centers and tiers.

### 3.3 Partitioning

Data will be highly partitioned, replicated, and distributed to meet the LSST requirements. Different partitions will be served by different disk arrays as well as different database servers. Data, or at least data organization and derived data products, might also be partitioned across data centers, for instance one center might specialize exclusively in temporal queries, while another will take care of certain type of detected objects.

Partitioning indexes is equally important—indices will be portioned into manageable chunks. Like ordinary data, indexes will be replicated and distributed.

### 3.4 Clustering

Co-locating data that is frequently accessed together, called *clustering,* can dramatically improve performance. Deciding how to cluster LSST data is particularly difficult, since data has both spatial and temporal locality, both very important. Replicating and efficiently clustering data for each dimension separately is not an obvious option, as it would immediately double the cost of storage. Instead, likely data will be clustered based on one dimension, and appropriate indexes will be built for the others. We are in the process of understanding the use cases and common queries that will imply what clustering algorithm is most effective for the LSST query load.

### 3.5 Replication and load balancing

As described above, data will be partitioned and spread across multiple disks arrays, servers or even centers. Based on our experience with production systems, simple replication will not be sufficient, as data is often accessed non-uniformly. Some parts of data are 'hotter' than others, and the hotspots move. This tends to overload some servers and leave others idle. We plan to auto-detect hot-spots and automatically rebalance and replicate data.

### 3.6 Provenance and virtual data

Some derived data products may be very large and may be relatively inexpensive to re-compute. If these products are seldom used, it is more effective to re-build them on the fly when requested rather than storing them. In order to support such *virtual data*, detailed information how to reproduce the data (i.e., *provenance*) will be kept. LSST data provenance includes a detailed record of how the data can be regenerated, in particular what software and parameters were used to produce the original version.

### 3.7 Data immutability, versioning, and backups

To satisfy data-reproducibility requirement, all production data released to users will be immutable (read-only). This greatly simplifies maintenance and improves manageability, as we no longer need to worry about files being changed, or having to re-synchronize existing file replicas. It also simplifies data backup.

Even though data is immutable, some changes are unavoidable. Inevitably, problems are discovered with the instruments or software, and as the experiment evolves better algorithms that might, for instance, better classify some objects are used to reprocess the original data. Versioning is used to satisfy both the reproducibility and 'update' requirements. Data is never changed, but released data will be corrected and improved through *versioning*. Instead of updating a released object, a new version will be created with the appropriate metadata. The system for tracking versions will allow access to either version, depending on user's needs. We have not yet determined the right versioning granularity. While it would be advantageous to version every object, for instance by time-stamping it, versioning all objects has non-negligible size (cost) implications.

We expect that both image data and databases will be fully replicated at data centers for load balancing and recovery in case the archive center suffers a disaster that cannot be recovered.

### 3.8 Data servers

The sheer scale of the LSST will require the deployment of hundreds if not thousands of database servers. We expect to run several independent server farms, each dedicated to serving a specific task or a specific user, and tuned accordingly for the expected query load.

As an example, SDSS observed that most "public" queries are short, but the volume of these queries is very high. The queries are also unpredictable, and often repeat (for instance students running the same query as part of their homework). At the same time, queries run by amateur and professional astronomers tend to run longer and be more complex. Likely two separate pools of servers, potentially at different centers will serve these two distinct groups of users.

### 3.9 Data access philosophy

Before being published, data must first pass multiple quality assurance (QA) tests – and in some cases peer-review tests. Many QA-tests take a long time and may require human evaluation. Peer review may take weeks. Periodic well-planned data releases allow this kind of quality control. On the other hand, some data, especially those related to generated transient alerts will be needed immediately to allow astronomers to study the cause of triggered alerts. This almost-real-time requirement leaves very little time for comprehensive validation of the data.

To address both requirements, two sets of data will be available to users: one set containing the "latest-greatest" data but use-at-own-risk type, and the second set that is well-curated, consistently processed, and peer-reviewed. We expect peer-reviewed versions about twice a year. Users may want to look at the "latest-greatest" data, but they should validate their results using the well QA'd, released data. We expect to have more freedom to refine "latest-greatest" data versions as the QA process uncovers problems, while the released version will all be consistently processed with the best current algorithms.

To allow cross-correlating LSST data with data from other surveys, the access to data will be provided through Virtual Observatory (VO)-compliant services. The LSST project is working with the National Virtual Observatory project [10] to ensure that these services are specified in a manner compatible with LSST-scale data volumes..

## 4. DATA INGEST

In LSST, ingest includes loading data into database, clustering (if necessary), indexing it, validating, building some derived datasets, making data immutable, backing up and replicating. The goal is to prepare the data for use by scientists to analyze the new data and compare it with historical data

LSST's scale and real-time requirements require automatic and efficient ingest. The ingest rate is 35 MB/sec or 1TB in a 10-hour observing window. This is not very demanding. However, several 'complications' destroy the simplicity of ingest. First, the real-time requirement requires data be loaded as it arrives. Even few seconds delay is unacceptable. Second, just-loaded data has to be immediately available for querying by other production processes (Association Pipeline[6]). And finally, by far the most challenging task is indexing the data: in order for the new data to become useful, it has to be incorporated into petabytes of existing data.

The real-time response is most important at the Base Facility, where alerts are generated. Luckily, we do not have to incorporate new data into petabytes of existing data there. Also the ingest data is easily parallelizable: per CCD [11]. At the Archive Center there is no real-time requirement, data ingest can take longer (up to 24 hours to load one-night worth of data if we do not want to fall behind), but it has to be merged with the existing "petabytes-of-data".

The current prototype that seems to meet both the Base Facility and the Archive Center ingest requirements involves loading data in parallel into some 200 independent *ingest tables* (one per CCD). Input data will be validated before the actual load. Our current model is to pre-load input data to an empty, temporary in-memory table and run queries to verify its correctness. Such an approach can isolate potential problems and quickly fix them without ever affecting the 'real' tables.

---

[6] The main task of the Association Pipeline is to associate sources from different times, filters, and perhaps sky positions, into data structures that describe astrophysical objects. It is doing that by reading data loaded into database by the previous stage: the *Detection Pipeline*.

The ingest tables will be truncated each night, which means they will never grow big enough to become a bottleneck for re-indexing. They will be directly queried by the Association Pipeline. Based on first tests, we should be able to easily ingest and re-index data for one image in less than 2-3 sec, which leaves plenty of time for processing before data from the next image arrives 12-13 sec later. To deal with failures, each image will be processed by two independent servers so that if one fails, the data will still be available from the other.

The ingest tables at the Base Facility are only needed to generate alerts[7], thus there is no reason to preserve the data. At the Archive Center, before truncating the ingest tables, data will be *merged*; i.e., ingested data will be added into the final, large tables. Merging is also a good opportunity to appropriately cluster data. Merging will collapse many small tables into a manageable number of much larger tables.

It is likely that a non-transactional engine will be used for loading data into staging tables before it is added to the catalogs. The main reason is to avoid the non-negligible transaction overhead. We can mainly afford that because the proposed architecture involves many small, tightly-controlled tables, so we can cleanly and quickly recover from potential failures by simply reloading data from the cached input files.

We still need to determine the optimal size of database files: files too small would increase metadata overhead, while files too large might degrade performance. It is expected database files will be somewhere between a few gigabytes and a few terabytes, (this range definitely needs to be narrowed down!).

The databases and tables used for ingesting data will run in a fully-controlled, production environment and will never be exposed to users. Query access to the LSST data will be provided through a dedicated set of databases.

## 5. APPROACH TO DESIGNING AND BUILDING THE SYSTEM

Designing the database system requires understanding the expected load, including common queries and data access patterns. It is impossible to precisely predict the load because of the LSST is unique: data will be made publicly available with minimal delays, and data will have both spatial and temporal aspects. But, we must try predicting the load, so we are examining and learning from the load on existing and past surveys.

### 5.1 Deadlines

The LSST camera will start taking data in 2012. The construction phase starts in 2009. We expect to make a final selection of the database system just before the construction phase starts. A preliminary choice, primarily for sizing the system, defining the baseline and prototyping will be made in second half of 2006.

### 5.2 UML-based approach

The database design is carried out using Unified Modeling Language (UML). Figure 1, for example, is a high-level snapshot of the *Run Catalog Ingest Service* use case, the service responsible for ingesting data into the Catalog Database.

One of many benefits of the UML-based approach is the ability to capture the design independently of the actual implementation. This is particularly important at this early stage, given how much can, and will change between now and the time we start constructing the production version of the system. UML lets us defer implementation details and keeps the model, including the SQL, "vendor independent". Since the database vendor-specific 'tricks' and extensions are the key to getting most out of the system, we are not avoiding them, just deferring them, to simplify a potential switch from one database system to another.

---

[7] To generate an alert, in addition to the object data from the current image, historical object data for that region of the sky is needed as well.

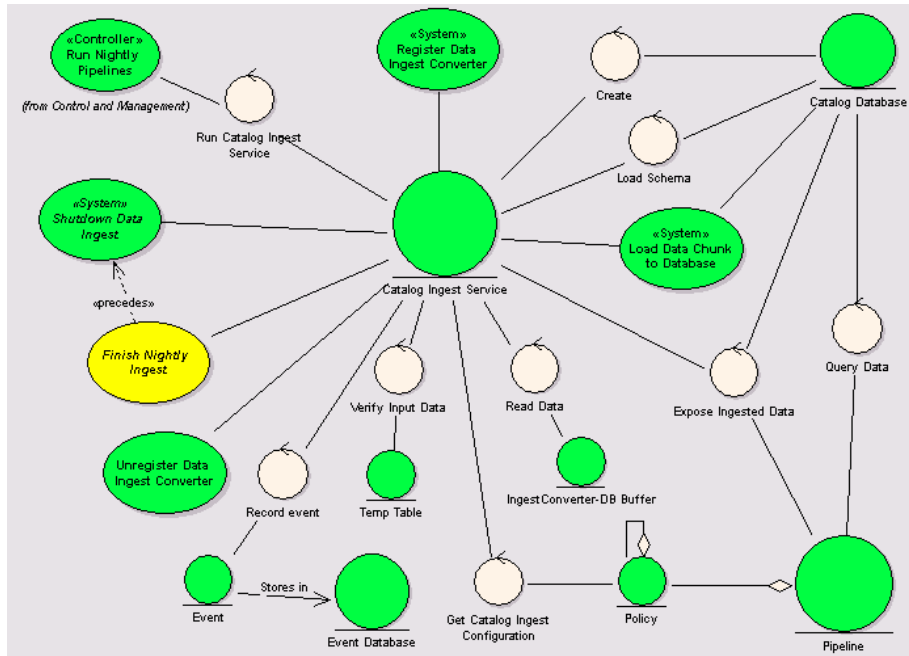

Figure 1: UML Robustness Diagram of the *Run Catalog Ingest Service* use case

**5.3 Database technology choice**

Our database design assumes RDBMS technology. While an ODBMS would be a more natural fit to the LSST needs, it is too risky to base a system that has to run through 2020 and beyond on a technology that is not widely accepted and used.

The LSST database system must support multi-petabyte scalability, good transactional insert performance, ad-hoc queries with an ability to scan data at tens of gigabytes per second, and high data availability -- all that at a reasonable cost. The largest production database systems in existence today are a few terabytes in size. These state-of-the-art systems rely on expensive hardware and on multiple copies of the data for fault tolerance, driving up the overall cost of such systems, not to mention the cost of the database software itself. A simple scale-up of such an approach is unlikely to be affordable for LSST.

The most promising systems that have recently emerged are open source relational databases. Their popularity, wide acceptance and performance, which often match the performance of their 'expensive' alternatives, only confirm they might be a viable option. Using an open source database also softens the issue of long-term support.

Current baseline, prototyping and bulk testing is done using MySQL. MySQL is one of the fastest database systems currently available and it offers a wide range of advanced features. Additionally, one of the next releases[8] will offer features geared towards very large database support, like a partitioning engine (which will allow systems to automatically partition data and queries), and the ability to store indexes for a single table in separate files. Another attractive feature of MySQL is an architecture that accommodates custom storage engines in place of the standard ones. This opens a wide range of possibilities, should we need to tailor the database specifically to our needs. It is a very common technique to customize the underlying database engine as much as possible for systems managing very large data sets, where every byte and every operation can visibly change the overall cost of the system, or its performance.

We strongly believe that close collaboration with the key developers of the underlying database technology is the key to success. Linking the developers of the database system with a real-world system that pushes their system to new limits is a powerful combination, as our past experience with Objectivity Inc. demonstrates. We are already closely collaborating with the MySQL developers to develop tools and techniques for supporting petascale databases.

---

[8] Most likely in release 5.2, which will be available in the next few months.

However, as there are counter-arguments in favor of commercial DBMS, in parallel, we are evaluating Microsoft's SQL Server and are planning to evaluate IBM's DB2 and expect to conduct a vendor challenge involving all of these DBMSes.

### 5.4 Prototyping

Constructing the production system is still several years away, so current effort is devoted to prototyping and testing. Tests include understanding physical limitations of the tested databases, uncovering potential bottlenecks and pitfalls, understanding feature sets, and understanding needed improvements. We are not focusing on low-level, fine-grain performance tuning, since such things will change in 5 years. The tests are done using a dedicated, relatively modest testbed. Managing a full-blown large-scale testbed would not be justified at this early stage of design.

### 5.5 Performance and scaling

This section briefly summarizes DBMS features needed to build a scalable system that performs well and meets the LSST requirements. Most were already mentioned; however it is useful to put them together in one place. They include:

- Good query optimization for ad hoc queries
- Complete suite of tools to administer data including backup, restore, reorganize, consistency check.
- Data recovery from errors at the page, file and table level.
- parallelization: data parallelization, index parallelization, query parallelization
- distributing data across multiple disk arrays, multiple database servers and multiple sites – both partitioning and replication
- support for files integrated with the database system
- co-locating (clustering) together data frequently accessed together
- tailoring the underlying database to meet specific LSST needs
- avoiding overhead of transactions when applicable
- clustering indexes: keeping indexes for the same table in the different files

The application (non-database) features include:

- automatic load balancing to avoid hot spots and replicating most used data
- executing queries at "the best place"
- making data immutable
- choosing "the right" database file size

The latest techniques to address scalability problems of today's large databases include data partitioning, index partitioning, replication, and query parallelization. While database systems have long supported these features, petabyte-scale databases required substantial improvements. It is very likely such features will be widely available, robust and will scale to LSST-level by 2012. If not, our backup plan is to come up with an in-house, customized solution, as we have done in the past for the petabyte BaBar system. With MySQL, such customization could be developed as a new storage engine, which could use the *Petacache* [5] developed at SLAC.

## 6. SUMMARY

The LSST will start collecting data in 2012 and is expected to reach multi-petabyte scale in the first year of operation. Building a multi-petabyte database at a reasonable cost is 'doable' today, but it requires significant effort. Well known techniques can not be simply scaled-up. The main challenges include the large data volume, the real-time requirements, the spatio-temporal aspects of the data, and making the data publicly available with minimal delays. To address these issues, LSST Data Management will rely on the concept of a hybrid system, with image data stored as files and "all the rest" managed by a database management system. The system will be highly distributed. We expect to replicate frequently accessed data for improved performance and reliability. To make a processing-storage tradeoff, some data will be virtual: it will be re-computed on demand. All released data will be immutable, and any updates will be handled by versioning. Several independent servers-farms will be tuned for particular group of users and their access patterns.

Ingesting data into the database will be done in parallel into short-lived ingest staging databases to improve performance. Data from the staging databases will be merged with the large production tables as soon as the data has been validated.

To capture the design independently of the actual database implementation, we use UML modeling and defer the implementation details until the last possible moment. Such an approach simplifies switching database implementations, should this be needed.

The current baseline and prototyping is done with MySQL, a promising open source, relational database. At the same time we are evaluating Microsoft's SQL Server and will soon look into IBM's DB2. The most attractive features of MySQL include a partitioning engine and pluggable architecture that allows one to easily replace storage engines with custom engines.

We are optimistic the production system will scale to required levels. Most likely an off-the-shelf solution will be used to provide partitioning, parallelization, and other advanced database features. Should such a system not be available, we will develop an in-house solution based on our experience with building other data-intensive applications.

## ACKNOWLEDGEMENTS

The LSST research and development effort is funded in part by the National Science Foundation under Scientific Program Order No. 9 (AST-0551161) through Cooperative Agreement AST-0132798. Additional funding comes from private donations, in-kind support at Department of Energy laboratories and other LSSTC Institutional Members.